# Simultaneous measurement of pressure evolution of crystal structure and superconductivity in FeSe$_{0.92}$ using designer diamonds

Walter Uhoya[1], Georgiy Tsoi[1], Yogesh Vohra[1], Nathaniel Wolanyk[2], Sistla Muralidhara Rao[3], Maw-Kuen Wu[3] and Samuel Weir[4]

[1] *Department of Physics, University of Alabama at Birmingham (UAB) - Birmingham, AL 35294, USA*
[2] *Illinois Wesleyan University (IWU) - Bloomington, IL 61702 - 2900, USA*
[3] *Institute of Physics, Academia Sinica - Nankang, Taipei 115, Taiwan*
[4] *Mail Stop L-041, Lawrence Livermore National Laboratory (LLNL) - Livermore, CA 94550, USA*

**Abstract** –Simultaneous high pressure x-ray diffraction and electrical resistance measurements have been carried out on a PbO type α-FeSe$_{0.92}$ compound to a pressure of 44 GPa and temperatures down to 4 K using designer diamond anvils at synchrotron source. At ambient temperature, a structural phase transition from a tetragonal (*P4/nmm*) phase to an orthorhombic (*Pbnm*) phase is observed at 11 GPa and the *Pbnm* phase persists up to 74 GPa. The superconducting transition temperature ($T_C$) increases rapidly with pressure reaching a maximum of ~28 K at ~ 6 GPa and decreases at higher pressures, disappearing completely at 14.6 GPa. Simultaneous pressure-dependent x-ray diffraction and resistance measurements at low temperatures show superconductivity only in a low pressure orthorhombic (*Cmma*) phase of the α-FeSe$_{0.92}$. Upon increasing pressure at 10 K near $T_C$, crystalline phases change from a mixture of orthorhombic *(Cmma)* and hexagonal *(P63/mmc)* to a high pressure orthorhombic *(Pbnm)* phase near 6.4 GPa where $T_C$ is maximum.

**Introduction.** – The pressure variable has always played a pivotal role in the discovery and optimization of novel superconducting materials. Discovery of high temperature superconductivity in a new class of iron-based layered compounds has received extensive attention recently [1-6]. Undoped iron-based layered compounds like REOFeAs (RE = trivalent rare earth metal), and AFe$_2$As$_2$ (A = divalent alkaline earth metal) are non-superconducting at ambient pressure and are known to exhibit tetragonal to orthorhombic structural transition and antiferromagnetic (AFM) ordering on cooling. The AFM ordering and structural transition is suppressed under high pressure or chemical doping and superconductivity is induced [1 - 4]. However, the critical relationships between structure, magnetism, and superconductivity still remain unresolved. More recently, superconductivity was reported at 8 K in α-FeSe$_{1-δ}$ samples with PbO-type tetragonal structure [5]. At ambient conditions, α-FeSe$_{1-δ}$ has a structure composed of stacks of edge-sharing FeSe$_4$-tetrahedral layers stacked along *c*-axis [5-7] while, the structure of FeAs-based superconductors consists of edge sharing FeAs$_4$-tetrahedra stacked layer by layer with separating elements like REO in REOFeAs or A in AFe$_2$As$_2$ between the FeAs$_4$ layers [1, 2]. The tetragonal α-FeSe undergoes a structural phase transition to an orthorhombic (*Cmma*) below 70 K upon cooling [7]. One remarkable aspect of superconductivity in binary FeSe-system is the strong relationship between the superconducting state and pressure. Recently, the $T_C$ onset was shown to increase at huge rate of dTc/dP = 9.1 K/GPa (the largest for any of the known FeAs-compounds), dramatically reaching an onset of 27 K at 1.5 GPa [6]. This sensitivity of $T_C$ to pressure convincingly indicates that there is a strong correlation between the superconducting properties and changes in the crystal structure of FeSe-system under pressure. A number of pressure-dependent structural and resistance measurements of the layered Fe-based systems have been reported that are aimed at understanding the critical relationship between compression behavior of crystal structure and superconductivity [6-11]. However, none of the previous works reported simultaneous high pressure and low temperature resistivity and x-ray diffraction measurements on the same sample in the same experiment. Here, we report simultaneous high pressure resistance and x-ray diffraction experiments using a designer-DAC to precisely elucidate the effect of pressure on the observed superconducting properties and the local crystallographic modulations of FeSe$_{1-δ}$. In addition, there have been some disagreements on structural phase modulations in FeSe$_{1-δ}$ system under pressure. While some report a tetragonal to a hexagonal phase (*P63/mmc*) transition above 12 GPa [8, 9], a phase transition from the tetragonal to an orthorhombic phase (*Pbnm*) has been reported by others [10, 11]. In addition, at the time of this study, reports for pressures above 40 GPa are not available. Thus, high pressure structural phases of FeSe$_{1-\delta}$ systems still need clarification and this is the second motivation of the present experiments. We thus present the results of our high pressure x-ray diffraction studies on FeSe system up to a high pressure maximum of 74 GPa at room temperature (RT) and 44 GPa at low temperatures (LT).

Table 1: Selected crystallographic parameters: lattice parameters (*a*, *b* , *c* and unit cell volume *V*), atom positions (x, y, z), percentage phase fraction ( Frac), Se:Fe occupancy ratio and goodness of fit ($x^2$) for different space groups (SG) of α-FeSe$_{0.92\pm0.04}$ sample obtained after Rietveld refinements of synchrotron XRD data collected at various temperatures and high pressures (P).

| P (GPa) | Ambient Temperature | | | Low Temperature of 4 K | | Low Temperature of 10 K | | |
|---|---|---|---|---|---|---|---|---|
| | 0.2 GPa | 0.2 GPa | 30 GPa | 4 GPa | 4 GPa | 12.9 GPa | 12.9 GPa | 24.3 GPa |
| SG | *P4/nmm* | *P63/mmc* | *Pbnm* | *Cmma* | *P63/mmc* | *Cmma* | *Pbnm* | *Pbnm* |
| a(Å) | 3.7687(2) | 3.6161(17) | 5.7142(2) | 5.1944(87) | 3.6627(13) | 4.7510(24) | 6.0686(24) | 4.2400(38) |
| b(Å) | 3.7687(2) | 3.6161(17) | 5.2034(9) | 5.3374(96) | 3.6627(13) | 5.2051(49) | 5.3745(20) | 5.6279(25) |
| c(Å) | 5.5073(6) | 5.8888(72) | 3.4168(8) | 5.1875(40) | 5.2173(13) | 4.4794(66) | 3.5368(12) | 3.9733(16) |
| V(Å$^3$) | 78.22 | 66.68 | 101.6 | 143.82 | 60.62 | 110.77 | 115.36 | 94.82 |
| Frac(%) | 91.06 | 8.94 | 100 | 60.6 | 39.4 | 16.1 | 83.3 | 100 |
| Fe(x, y, z) | 0.75, 0.25, 0 | 0 , 0, 0 | 0.232(3),0.534(3),0.25 | 0.25, 0, 0 | 0 , 0, 0 | 0.25, 0, 0 | 0.2578, 0.5051, 0.25 | 0.4701, 0.605, 0.25 |
| Se(x, y, z) | 0.25,0.25, 0.2902(3) | 0.3333, 0.6667,0.25 | 0.909(7),0.773(8),0.25 | 0, 0.25,0.7371 | 0.3333, 0.6667, 0.25 | 0, 0.25, 0.6717 | 0.8983, 0.7495, 0.25 | 0.9695,0.8389, 0.25 |
| Se : Fe | 0.913 | 1 | 1 | 0.9228 | 1 | 1 | 0.965 | 1 |
| x$^2$ | 0.1 | 0.1 | 0.1238 | 0.4407 | 0.4497 | 0.1768 | 0.1768 | 0.1 |

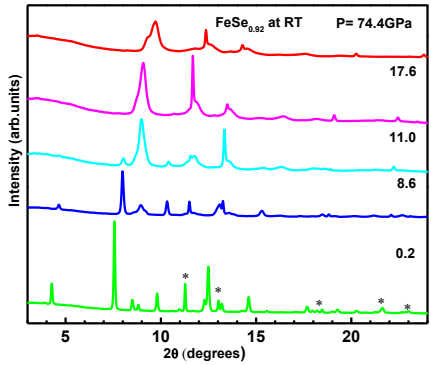

Fig. 1: (Color on-line) XRD profiles for FeSe$_{0.92}$ at various pressures up to 74 GPa and room temperature (RT). A structural transition from a tetragonal *(P4/nmm)* phase to an orthorhombic (*Pbnm*) phase occurs at ~11 GPa and the *Pbnm* phase persists up to 74.4 GPa. The diffraction peaks from fcc phase of Cu pressure marker are marked by asterisk (*).

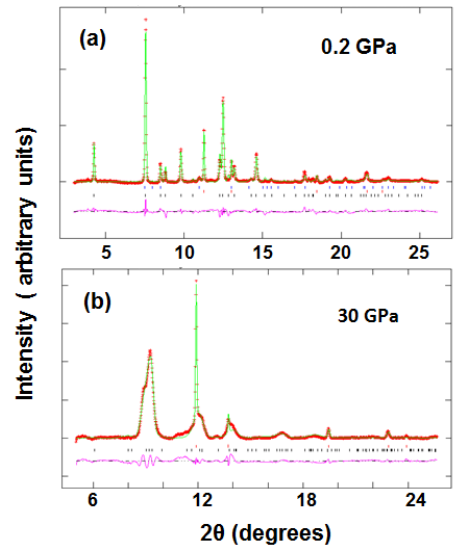

Fig. 2: (Colour on-line) Refined XRD patterns of FeSe$_{0.92}$ at ambient temperature. The lowermost solid line (magenta) in (a) and (b) are the difference profile curves between the observed (red marker) and calculated (green line) profiles. The pattern at 0.2 GPa comprises of primary tetragonal (*P4/nmm*) phase (bottom tick marks, 91%), and a secondary hexagonal (*P63/mmc*) phase (top tick marks, 9%). The high pressure phase at 30 GPa is refined in the orthorhombic (*Pbnm*) phase (bottom tick marks). The reflections from Cu – pressure marker are denoted by middle tick marks at 0.2 GPa and top tick marks at 30 GPa.

**Experimental Details.** – Large single-crystalline of FeSe$_{1-\delta}$ samples were grown out of Fe and Se powders using convective solution transport technique with KCl solvent as described in the original publication [12]. These crystals were annealed *in situ* at 400-350$^o$C for 20-30 hours to improve their superconducting properties, and then cooled to RT. The crystals were characterized using high resolution transmission electron microscopy, energy dispersive x-ray analysis, powder x-ray diffraction (XRD), and Raman spectroscopy. The resistance measurements showed anomaly corresponding to tetragonal to orthorhombic transition at 90 K and superconducting transition, T$_C$ of 8.5 K at ambient pressure [12]. For high pressure studies, the crystals were ground and loaded into an 80 µm hole of a spring-steel gasket that was first pre-indented to a ~50-µm thickness and mounted between a matched pair of (300 µm size culet) beveled diamond anvil cell (DAC) ready for high-pressure XRD measurements. The synchrotron XRD experiments were performed at the high pressure beam-line 16-BM-D, APS at Argonne National Laboratory. An angle dispersive technique with a MAR345 image-plate area detector was employed using a focused monochromatic beam with x-ray wavelength, λ = 0.409892 Å and sample to detector distance of 320.5 mm. The image plate XRD patterns were recorded with a focused x-ray beam of 5μm by 5μm on an 80 μm diameter sample mixed with powdered copper pressure marker. Experimental geometric constraints and the sample-to-image plate detector distance were calibrated using CeO$_2$ diffraction pattern and were held at the standard throughout the entirety of the experiment. The lattice parameters of the Cu-pressure marker obtained from the Rietveld refinement of the XRD patterns of Cu-powder mixed with the sample in the DAC were employed for pressure calibration [13]. At ambient temperature, the Birch-Murnaghan equation [14] as shown by eq. 1 was fitted to the available equation of state data on Cu-pressure standard [13] to determine the sample pressure, where $B_0$ is the bulk modulus, $B_0'$ is the first derivative of bulk modulus at ambient pressure, and $V_0$ is the ambient pressure volume.

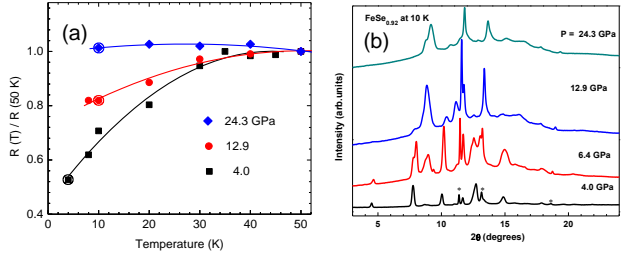

Fig. 3: (a) Temperature dependence of electrical resistance for FeSe$_{0.92}$ measured using designer-DAC at high pressures. XRD measurements were taken simultaneously with resistivity measurement at each data point. Continuous lines are guide to the eye. (b) XRD patterns of FeSe$_{0.92}$ at 10 K and various pressures up to 24.3 GPa showing a pressure induce *Cmma* to *Pbnm* structural transition. The reflections from *fcc* phase of Cu pressure marker are denoted with asterisks (*).

$$P = \tfrac{3}{2} B_0 [(\tfrac{V_0}{V})^{7/3} - (\tfrac{V_0}{V})^{5/3}]\{1 + \tfrac{3}{4}(B_0'-4)[(\tfrac{V_0}{V})^{2/3} - 1]\} \quad (1)$$

The high-pressure and low-temperature electrical resistance measurements were performed down to 10 K using four-probe method in an eight-probe designer DAC [15, 16] as described earlier [17]. Pressure was applied using a gas membrane DAC and monitored by the ruby fluorescence technique [17, 18]. For simultaneous resistivity and x-ray diffraction experiments, the designer DAC was cooled down in a continuous helium flow-type-cryostat, and the pressure in the cell was measured *in situ* with the ruby fluorescence technique [17, 18]. Powdered copper pressure standard mixed with the sample was employed as secondary pressure gauge for low temperature experiments. No pressure transmitting medium was employed. So the present studies can be regarded as nonhydrostatic where the uniaxial stress component is limited by the sample shear strength at high pressures. The software package FIT2D [19] was used to integrate the collected MAR345 image plate diffraction patterns which were analyzed by GSAS [20] software package with EXPGUI interface [21] employing full-pattern Rietveld refinements and Le Bail fit techniques to extract structural parameters.

**Results and Discussions.** – Figure 1 shows representative XRD patterns of α-FeSe$_{0.92}$ sample and Fcc-Cu pressure standard obtained at RT and various pressures up to 74 GPa. At RT and lowest pressure of 0.2 GPa, Rietveld refinement of XRD patterns revealed a tetragonal *P4/nmm* (α-FeSe) phase with a phase fraction of 91.1%, Fe:Se occupancy ration of 0.913, and lattice parameters $a = b = 3.7687(2)$ Å, $c = 5.5073(6)$ Å, $V = 78.221(8)$ Å$^3$ (8) as shown in Fig. 2(a) and Table 1. The α-FeSe$_{0.92}$ has the following crystalline arrangement: Fe atoms occupy the 2$a$ position (1/4, -1/4, 0) and Se atoms occupy the 2$c$ positions (1/4, 1/4, z) with refined z = 0.2902 at 0.2 GPa. Besides, we observed distinct diffraction peaks corresponding to a minority NiAs-type hexagonal phase (β-FeSe). At ambient conditions, the β-FeSe-phase was well refined with the lattice parameters $a = 3.6161(17)$ Å, $b = 3.6161(17)$ Å and $c = 5.8888(72)$ Å with a phase fraction of 8.94%. At room temperature, high pressure diffraction patterns from 11 GPa and above look remarkably different from the low pressure tetragonal *(P4/nmm)* patterns as shown in Fig. 1, suggesting an onset of a new pressure induced crystallographic phase. The phase above 11 GPa was well refined using an orthorhombic *Pbnm* model which is in agreement with earlier reports [10, 11] but in contrast to hexagonal phase reported in Refs. (8, 9) above 12 GPa (this may be a result of the larger volume of the tetragonal phase compared to the earlier reports). Figure 2(b) shows typical Rietveld refined high pressure phase of FeSe$_{0.92}$ using the orthorhombic *(Pbnm)* phase at 30 GPa and 300 K. The results of refinements and structural parameters are summarized in Table 1. The *Pbnm* phase was found to be stable up to our presume limit of 74 GPa with lattice constants: $a = 4.7657$Å, $b = 4.2290$Å, $c = 3.1455$Å and V = 63.4Å$^3$ at this pressure.

Figure 3(a) shows typical resistance data obtained from simultaneous electrical resistivity and XRD of α-FeSe$_{0.92}$ measured in situ at high pressures and temperatures down to 4 K. Representative pressure dependent XRD patterns obtained simultaneously with resistance measurement at 10 K and various pressures up to 24.3 GPa are presented in Fig. 3(b). The measured resistance shows a resistive anomaly characterized by a rapid down turn of R vs. T curve at low temperatures indicating an onset of superconductivity. The $T_c$ is determined by the intersection of the two linear fits to data above and below $T_C$, as illustrated in Fig. 5 for P = 4.4 GPa.

It has been shown that FeSe undergoes a tetragonal *(P4/nmm)* to an orthorhombic *(Cmma)* phase transition below 70 K [7], suggesting that $T_c$ occurs in the *Cmma* phase at low pressures. Figure 4 shows typical Rietveld refinement results of the XRD data obtained simultaneously with resistance measurements corresponding to the three encircled resistance data points depicted in Fig. 3(a). Figure 4(a) shows the observed, calculated and difference profiles obtained after combined Rietveld refinement of FeSe$_{0.92}$ in the *Cmma* and *P63/mmc* phases at 4 GPa and 4 K. The pattern is well refined in the *Cmma* phase with the parameters: $a$ = 5.1944(87) Å, $b$ = 5.3374(96) Å, $c$ = 5.1875(40) Å, V=143.824(87) Å$^3$ and phase fraction of 60.6%. Minority hexagonal β-FeSe phase observed at ambient pressure is also present in the low temperature structure with refined lattice parameters $a$ = 3.6627 and $c$ = 5.2173 and with phase fraction of 39.4% (see Table 1 for the refined structural parameters). Our structural results at low temperatures are consistent with earlier reports [7, 11].

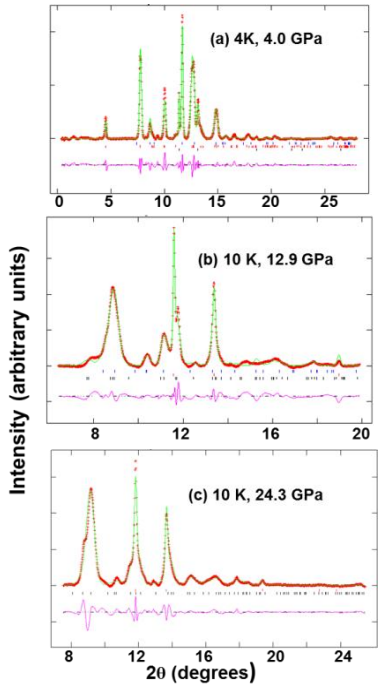

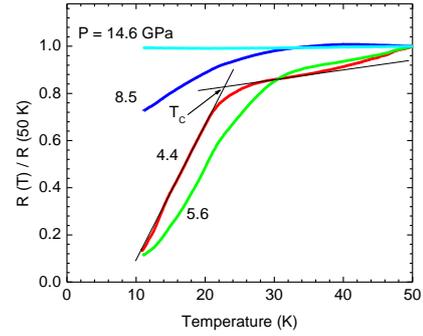

Fig. 5: (Color online) The measured four probe resistance of the α-FeSe$_{0.92}$ sample as a function of temperature at various pressures up to 14.6 GPa and down to 10 K. Determination of superconductivity (Tc) of 22.9 K is shown for P=4.4 GPa.

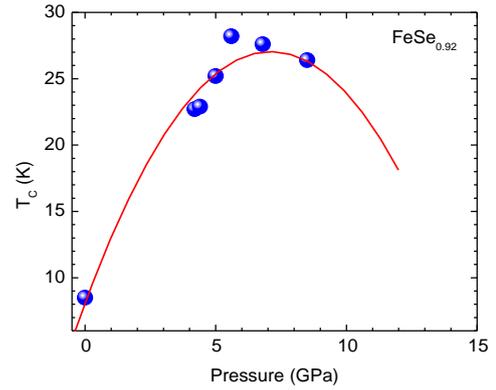

Fig. 6: Pressure dependence of the measured onset of superconducting transition temperature ($T_C$) for FeSe$_{0.92}$ measured using designer diamond anvil technology.

Fig. 4: (Colour on-line) Refined XRD spectrum of FeSe$_{0.92}$ in the low temperature orthorhombic phase (*Cmma*) and secondary hexagonal (*P63/mmc*) phase at low pressures and in the orthorhombic (*Pbnm*) at high pressures. The lowest solid line (magenta) in Fig. 4(a-c) are the difference profile curves between the observed (red data points) and calculated (green line) profiles. (a) Shows reflections corresponding to *Cmma* phase (middle tick marks), *P63/mmc* phase (top tick marks) and *Fcc*-Cu pressure marker (bottom tick marks); (b) shows *Cmma* phase (top tick marks), *Pbnm* (bottom tick marks), and *Fcc*-Cu (middle tick marks), and (c) shows Pbnm phase (bottom tick marks) and Fcc-Cu (top tick marks).

On application of higher pressure at low temperatures, the reflections in the diffraction profile between 6 and 12.9 GPa first develop a characteristic broadening, signifying the onset of a structural transition to a different phase, the peak at 4.5 degrees disappears and the high pressure phase looks much different above 12.9 GPa as shown in Fig. 3(b). The XRD patterns in the transition region between 6 and 12.9 GPa shown in Fig. 3(b) were well refined using a combination of both *Pbnm* and *Cmma* structure models. Combined Rietveld refinement of the pattern at 12.9 GPa and 10 K revealed a major *Pbnm* (83.3%) and small *Cmma* (16.1%) phases (Fig. 4(b) and Table 1). The phase fraction of *Cmma* deceases with pressure until ~ 24.3 GPa at 10 K where the sample is fully transformed to *Pbnm* phase (see Table 1) and this is consistent with earlier report which showed a mixed *Cmma* and *Pbnm* phases up to 26 GPa at 8 K [22]. Figure 4(c) shows typical GSAS refinement of the high pressure FeSe$_{0.92}$ in the *Pbnm* phase and *fcc*-Cu pressure marker at 24.3 GPa. The pattern is well refined without any indication of a secondary phase and the difference between the observed and the calculated pattern is quite satisfactory, suggesting that the high pressure structure is evidently orthorhombic *Pbnm*. The refined lattice constants obtained for the *Pbnm* phase at 24.3 GPa are $a$ = 4.2400(38) Å, $b$ = 5.6279(25) Å, $c$ = 3.9733(16) Å and V=94.82 Å$^3$. A comparison of the *Pbnm* structure at RT and 30 GPa (see Table 1 for lattice parameters) suggests that the low temperature phase at 10 K and 24.3 GPa is ~6 times denser than the RT phase at 30 GPa. This suggests a temperature induced structural transition to a denser *Pbnm* phase at lower temperatures. This is consistent with Ref. (22) where a *Pbnm* phase with V=105.3 Å$^3$ at RT and 19 GPa was observed as compared to a more dense low temperature *Pbnm* phase with V=99.9 (~5% denser) at 18 GPa and 8 K." Further work is needed to fully characterize this orthorhombic structural transition. The *Pbnm* phase observed here is stable at low temperatures up to our maximum pressure of 44 GPa at LT.

Figure 5 shows the measured four probe resistance of the α-FeSe$_{0.92}$ sample as a function of temperature at various pressures up to 14.6 GPa and down to 10 K. The resistance measurements were performed independent of structural measurement on a sample taken from the same lot as those used in simultaneous structural and resistivity experiments discussed in the previous section. A similar trend of temperature dependence of resistance for pressures less than 8.5 GPa is observed: the measured resistance shows a resistive anomaly characterized by a gradual down turn in R vs. T curve at low temperatures indicating an onset of superconductivity. The gradual resistive transition at high pressures is similar to broad resistive transition with zero resistance at 7.5 K with an onset at 11 K observed in similar FeSe$_{1-\delta}$ sample taken from the same batch as those used in our pressure studies [12]. The broad transition is probably connected to sample inhomogeneity resulting from presence of a second hexagonal phase reported in these crystals [6-12]. Finally, at 8.5 GPa, the superconducting transition signal broadens considerably, indicating a possible phase transformation to a non-superconducting state. However, an additional effect of non-hydrostatic pressure gradient which may cause additional broadening of superconducting transition cannot be ruled out. Further increase in pressure above 9 GPa leads to enhancement of resistance at low temperatures suggesting that the high pressure phase is a semiconducting phase and no traces of superconductivity are observed at 14.6 GPa. These observations are consistent with our XRD results which reveal a *Cmma* to *Pbnm* transition occurring over a broad pressure range from 6 GPa to 12.9 GPa. It is interesting to note that recent XRD and resistance measurement on Fe$_{1.01}$Se show that the tetragonal α-FeSe transforms to a denser β-FeSe which is nonmagnetic and semiconducting above 9 GPa [8].

Figure 6 shows the variation of T$_C$ as a function of pressure (P) which can be described by a polynomial regression given by eq. 2 where Tc is in K and pressure P, in GPa.

T$_C$ = (8.08±1.2) K + (5.33±0.43) KGPa$^{-1}$P- (0.37±0.04) KGPa$^{-2}$P$^2$       (2)

The measured Tc increases rapidly with increasing pressure to 6 GPa and then shows a gradual decrease beyond this pressure. The average value of $dTc/dP$ at zero pressure is 5.33 KGPa$^{-1}$. This is lower than the value of $dT_c/dP$ = 9.1 GPa$^{-1}$ reported earlier in measurements confined to the lower pressure region between 0 and 1.48 GPa [6]. The maximum T$_C$ onset from our measurement is 28 K at 6 GPa. Equation 2 suggests that T$_C$ onset disappears completely above 15.8 GPa. These observations coincide with our XRD studies where the α-FeSe$_{0.92}$ sample transforms from the low pressure *Cmma* phase to the high pressure *Pbnm* phase over a broad pressure range, starting at 6 GPa to 12.9 GPa. Sudden decrease of T$_C$ at the onset of *Pbnm* phase and its complete disappearance above 15.8 GPa, where the structure is fully transformed in to *Pbnm* phase, suggests that superconductivity is favored in the *Cmma* phase but not in the *Pbnm* phase. In addition, at RT and 11 GPa, α-FeSe$_{0.92}$ transforms from a tetragonal *(P4/nmm)* to an orthorhombic *(Pbnm)* and the orthorhombic phase remains stable to 74 GPa. The *P4/nmm* to *Pbnm* phase transition at 11 GPa is preceded by a rapid drop in the tetragonal unit cell volume and *c/a* ratio between 0.2 and 8.1 GPa, where T$_C$ is maximum. Recent XRD measurements on Fe$_{1.01}$Se at RT [8] and FeSe$_{1.03}$ at 16 K [11] and high pressures have also shown a strong interlayer compressibility of FeSe at high pressure that may cause a strong enhancement of T$_C$ in this material. Reported results from independent resistance measurements show that T$_c$ is enhanced with pressure and passes through a maximum between ~6 and ~9 GPa [8, 9, 11 ] consistent with our present simultaneous measurements.

**Conclusions.** – In summary, we have presented high pressure superconductivity data on the iron-based layered superconductor FeSe$_{0.92}$ to pressures up to 44 GPa and temperatures down to 4 K. The superconducting transition temperature increases rapidly with increasing pressure, reaching a maximum of 28 K at 6 GPa, and then shows a decrease beyond this pressure. We did not detect any evidence of T$_c$ above 14.6 GPa and down to 10 K in the *Pbnm* phase. This observation coincides with the room temperature pressure-dependent XRD results, indicating that the FeSe$_{0.92}$ sample undergoes a pressure-induced phase transformation from a tetragonal to an orthorhombic (*Pbnm*) phase at 11 GPa and the *Pbnm* phase remains stable up to the pressure limit of 74 GPa. Our simultaneous structural and transport studies revealed a *Cmma* phase at low temperatures and pressures where the sample is superconducting. The *Cmma* transforms gradually to an orthorhombic (*Pbnm*) at high pressures starting at ~ 6 GPa where Tc is maximum, and the *Pbnm* phase became a major phase above 12.9 GPa at 10 K. Our simultaneous resistivity and structural studies at low temperatures indicate that the orthorhombic (*Pbnm*) phase is non-superconducting.

∗∗∗

Nathaniel Wolanyk acknowledges support from the National Science Foundation (NSF) Research Experiences for Undergraduates (REU)-site under grant no. NSF-DMR-1058974 awarded to UAB. Walter Uhoya acknowledges support from the Carnegie/Department of Energy (DOE) Alliance Center (CDAC) under grant no. DE-FC52-08NA28554. Portions of this work were performed in a synchrotron facility at HPCAT (Sector 16), Advanced Photon Source (APS), Argonne National Laboratory.